\documentclass[aps,amsmath,notitlepage,twocolumn,amssymb,prl,longbibliography,showpacs]{revtex4-1}

\usepackage{graphicx}
\usepackage{dcolumn}
\usepackage{bm}
\usepackage[colorlinks=true,linkcolor = blue,citecolor=blue,urlcolor=blue,anchorcolor = blue]{hyperref}
\usepackage{color}
\usepackage{wasysym}
\usepackage{stmaryrd}
\usepackage{verbatim}
\usepackage{subfigure}
\usepackage{amsmath}
\usepackage{times}
\usepackage[version=4]{mhchem}
\usepackage{braket}
\usepackage{siunitx}
\usepackage{siunitx}
\usepackage{amssymb}
\newcommand{\RNum}[1]{\uppercase\expandafter{\romannumeral #1\relax}}

\begin{document}

\title{\large \bf Rare-earth chalcohalides: A family of van der Waals layered Kitaev spin liquid candidates}

\author{Jianting Ji$^{1}$}
\thanks{These authors contributed to the work equally.}

\author{Mengjie Sun$^{1,2}$}
\thanks{These authors contributed to the work equally.}

\author{Yanzhen Cai$^{3}$}

\author{Yimeng Wang$^{1,2}$}

\author{Yingqi Sun$^{1,2}$}

\author{Wei Ren$^{3}$}

\author{Zheng Zhang$^{1,2}$}

\author{Feng Jin$^{1}$}

\author{Qingming Zhang$^{3,1}$}
\email[e-mail:]{qmzhang@ruc.edu.cn}

\affiliation{$^{1}$Beijing National Laboratory for Condensed Matter Physics, Institute of Physics, Chinese Academy of Sciences, Beijing 100190, China}
\affiliation{$^{2}$Department of Physics, Renmin University of China, Beijing 100872, China}
\affiliation{$^{3}$School of Physical Science and Technology, Lanzhou University, Lanzhou 730000, China}

\date{\today}

\begin{abstract}
Kitaev spin liquid (KSL) system has attracted tremendous attention in past years because of its fundamental significance in condensed matter physics and promising applications in fault-tolerant topological quantum computation. Material realization of such a system remains a major challenge in the field due to the unusual configuration of anisotropic spin interactions, though great effort has been made before. Here we reveal that rare-earth chalcohalides REChX (RE=rare earth, Ch=O, S, Se, Te, X=F, Cl, Br, I) can serve as a family of KSL candidates. Most family members have the typical SmSI-type structure with a high symmetry of R-3m and rare-earth magnetic ions form an undistorted honeycomb lattice. The strong spin-orbit coupling of 4f electrons intrinsically offers anisotropic spin interactions as required by Kitaev model. We have grown the crystals of YbOCl and synthesized the polycrystals of SmSI, ErOF, HoOF and DyOF, and made careful structural characterizations. We carry out magnetic and heat capacity measurements down to 1.8 K and find no obvious magnetic transition in all the samples but DyOF. The van der Waals interlayer coupling highlights the true two-dimensionality of the family which is vital for the exact realization of Abelian/non-Abelian anyons, and the graphene-like feature will be a prominent advantage for developing miniaturized devices. The family is expected to act as an inspiring material platform for the exploration of KSL physics.
\end{abstract}


\maketitle

\section{INTRODUCTION}
Quantum spin liquid (QSL) refers to a novel quantum spin disordered state with strong spin entanglement, which hosts many exotic excitations and phenomena \cite{ANDERSON1973153}. Generally it can be driven by geometrical spin frustrations plus quantum spin fluctuations. Kitaev proposed an exactly solvable spin model on honeycomb lattice with the frustration of exchange coupling rather than a geometrical one \cite{KITAEV20062}, which offers an alternative way to realize QSL. The most intriguing feature of the model is that it has a gapless QSL phase which allows excitations of non-Abelian anyons when a spin gap is stabilized by external magnetic fields. This opens the door to the scheme of topological quantum computation based on Kitaev spin liquid.
\par

On the other hand, the bond-dependent anisotropic exchange coupling is really unusual and challenges general experience in material realization of quantum spin systems. Strong spin-orbit coupling (SOC) is believed to be a feasible mechanism yielding such kind of anisotropic spin interactions\cite{PhysRevB.78.094403}. Jackeli and Khaliulin proposed an attractive scenario based on SOC \cite{PhysRevLett.102.017205}, in which both spin-orbit-coupled effective moments and “spin”interactions between them are highly anisotropic. The isotropic term can be canceled out through a delicate arrangement of mediating anions and the bond-dependent term becomes dominant. This greatly stimulates and clearly guides the exploration of 4d/5d Kitaev materials.
\par
The 5d iridates A$_2$IrO$_3$ (A=Li, Na, Cu) are the first Kitaev materials predicted by Jackeli-Khaliulin mechanism \cite{PhysRevLett.102.017205,Kobayashi2003,PhysRevB.82.064412,Abramchuk2017}. The Ir$^{4+}$ (5d$^5$) ions are centered in the edge-sharing IrO$_6$ octahedra and form a honeycomb network normal to the diagonals. A crystalline electric field (CEF) ground state of Kramers doublet with an effective spin J$_{eff}$=1/2, is created in the ions by the strong SOC and the CEF effect. The iridates are magnetically ordered at low temperatures, indicating a non-KSL ground state. The related compound A$_3$LiIr$_2$O$_6$ (A=Ag, Cu, H), which is structurally similar to the original iridates, was also reported \cite{TODOROVA20111112,Roudebush2016,Kitagawa2018}. Particularly in H$_3$LiIr$_2$O$_6$, no magnetic transition was observed down to 50 mK. Whether or not this is an evidence for KSL ground state, still remains to be clarified, because of the substantial existence of quenched disorders in the compound \cite{Kitagawa2018}. The 4d$^5$ Li$_2$RhO$_3$ with simlar structure and J$_{eff}$=1/2  was also investigated and a partial spin-freezing was observed at 6 K \cite{PhysRevB.96.094432}.
\par

Recently, the 4d ruthenium halide $\alpha$-RuCl$_3$ was extensively studied as a KSL candidate \cite{KOBAYASHI1992,PhysRevB.90.041112,PhysRevB.91.144420,PhysRevB.91.180401,PhysRevB.93.134423,PhysRevB.96.064430,Banerjee2017,Kasahara2018}. Similar to the Ir$^{4+}$ ions in the iridates, Ru$^{3+}$(4d$^5$) ions have an effective spin J$_{eff}$=1/2 from Kramers doublet guaranteed by the strong SOC and the CEF effect, but sit on a slightly distorted honeycomb lattice at room temperature. It undergoes an antiferromagnetic (AFM) transition around 8 K, also indicating a non-KSL ground state. Neutron scattering experiments reported the unusual signals possibly from Majorana excitations, suggesting $\alpha$-RuCl$_3$ is proximate to a KSL system \cite{Banerjee2017}. With the suppression of the magnetic transition by in-plane fields, Kasahara et al. reported the observation of a half-integer quantum Hall effect, which was considered as an indication of Majorana excitations in thermodynamic channel \cite{Kasahara2018}.
\par

\begin{figure}
\centering
\includegraphics[width=\linewidth]{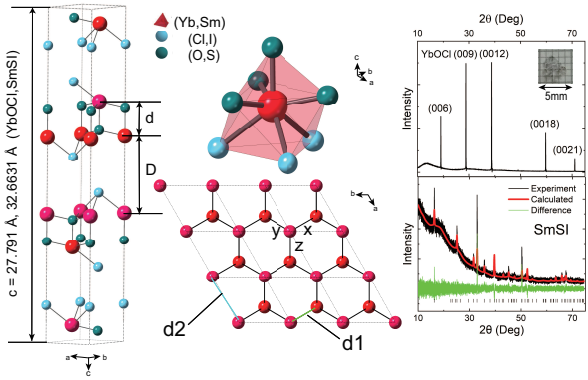}
\caption{Crystal structure of YbOCl/SmSI and their structural refinements. The upper middle shows the polyhedron surrounding Yb/Sm and the lower middle is the top view along c axis showing the honeycomb lattice formed by Yb/Sm ions. The elevated XRD background at small angles in SmSI comes from the sticky tape which was used to protect the air-sensitive sample.}
\label{Fig.1}
\end{figure}
\par

Along the way of trichlorides, more KSL candidates have been explored very recently. YbCl$_3$, having a monoclinic structure and an effective spin J$_{eff}$=1/2, was suggested to be proximate to the KSL physics \cite{PhysRevB.102.014427}. And the short- and long-range magnetic transitions were observed at low temperatures in the compound. The 5d osmium trichloride Os$_x$Cl$_3$ was reported to be a KSL candidate very recently \cite{Kataoka2020}. Just like the cases mentioned above, Os$^{3+}$ (5d$^5$) ions have an effective spin J$_{eff}$=1/2, but are originally sit on a triangular rather than honeycomb lattice. It was suggested that there exist nano-domains in which Os$^{3+}$ ions are placed on a honeycomb network. The deficiency of osmium remains a crucial issue in the material.
\par

As we can see, the current effort for searching for KSL candidates mainly focuses on 4d/5d systems, as inspired by the original Jackeli-Khaliulin scenario. Actually, most rare-earth 4f ions typically have strong SOC and an effective spin 1/2 guaranteed by Kramers doublet, and hence are natural building blocks for KSL candidates. Theoretical considerations in this aspect have been made \cite{PhysRevB.95.085132,PhysRevB.98.134437,10.21468/SciPostPhysCore.3.1.004}, while rare-earth-based KSL materials are little explored at present. In our previous studies, we have revealed the geometrically frustrated rare-earth-based quantum magnet YbMgGaO$_4$ and the family of rare-earth chalcogenides \cite{Li2015,PhysRevLett.115.167203,Liu_2018}, which serve as an ideal platform for the exploration of QSL physics. In the rare-earth materials, exchange coupling is relatively small (normally several kelvins) because of the screening of the outer 5s/5p shells. But this also brings us some advantages, for example, one can safely consider the nearest-neighbor spin interactions without worrying about further neighbors and the low-energy spin Hamiltonian can be well investigated by the application of magnetic fields generated in the lab.
\par
\begin{figure}
\centering
\includegraphics[width=\linewidth]{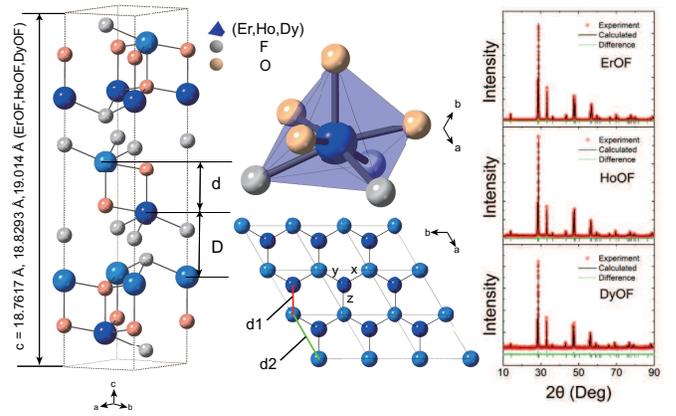}
\caption{Crystal structure of ErOF/HoOF/DyOF and their structural refinements. The upper middle shows the polyhedron surrounding Er/Ho/Dy and the lower middle is the top view along c axis showing the honeycomb lattice formed by Er/Ho/Dy ions.}
\label{Fig.2}
\end{figure}
\par

In this paper, we point out that rare-earth chalcohalides REChX (RE=rare earth, Ch=O, S, Se, Te, X=F, Cl, Br, I) may be a promising family of KSL materials. The general crystal structure of the family belongs to the SmSI-type or its polytypes which basically show a high symmetry of R-3m. This allows rare-earth magnetic ions to form an undistorted honeycomb lattice. The honeycomb magnetic layers are well separated and there is only a weak van der Waals coupling between the layers. We have grown the crystals of YbOCl and synthesized the polycrystals of SmSI, ErOF, HoOF and DyOF. They are structurally determined by X-ray diffraction measurements and careful Rietveld refinements. We carry out magnetic susceptibility and heat capacity measurements down to 1.8 K, which show no obvious phase transitions in the samples. We further discuss anisotropic spin interactions in the compounds.
\par

\section{Synthesis of crystals/polycrystals and experiments}
The crystals of YbOCl have been grown by a standard flux method \cite{Brandt1974}. The operations related to raw materials are carried out in a glove box. Yb$_2$O$_3$ (99.9$\%$, Alfa) and anhydrous YbCl$_3$ (99.9$\%$, Alfa) are mixed with the ratio of 1:10, and the mixed raw materials are sealed in a quartz tube filled with argon gas. The sealed tubes are placed in a box furnace and heat up to 1050 $^{\circ}$C at which the furnace stay for 4 days. Then the furnace is cooled down to room temperature with the rate of 8 $^{\circ}$C per hour. The products are filtered and cleaned in distilled water. Finally the nearly colorless YbOCl crystals are obtained with the maximum size of $\sim$5.0x5.0x0.1 mm$^3$ (see the inset in Fig. \ref{Fig.1}).
\par

\begin{figure}
\centering
\includegraphics[width=\linewidth]{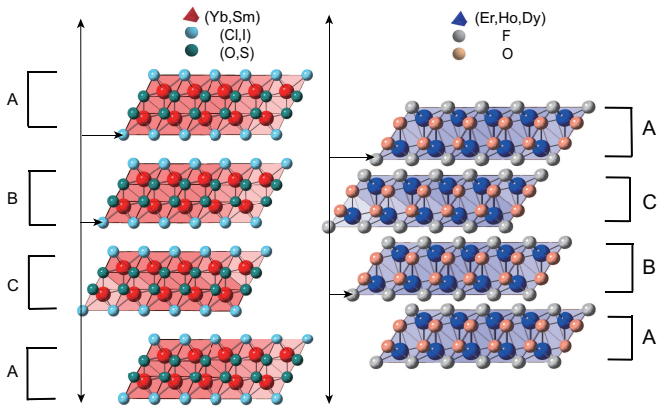}
\caption{Different stacking order in the compounds studied here. Adjacent magnetic layers form charge-neutral double-layer sheets which are well separated with a weak van der Waals interaction. YbOCl/SmSI has an ABC stacking order (left panel) while an ACB order for ErOF/HoOF/DyOF (right panel).}
\label{Fig.3}
\end{figure}
\par

The SmSI polycrystals are synthesized with a standard solid state reaction method \cite{Savigny1973,Beck1986}. Samarium and sulfur powder together with iodine balls, are mixed up with the ratio of 1:1:1 in a glove box and sealed in a quartz tube. The mixtures are placed in a box furnace and heated up to 450 $\sim$ 500 $^{\circ}$C at which the furnace stays for some time. Then temperature is slowly increased to 830 $\sim$ 900 $^{\circ}$C at which the furnace stays for one week. The polycrystals are obtained after the furnace is naturally cooled down to room temperature. It should be noted here that SmSI is quite unstable in air. The sample is sealed with sticky tape when making structural characterizations.
\par
The polycrystals of ErOF, DyOF and HoOF are also synthesized by a solid state reaction method \cite{Podberezskaya1966}. The raw materials are Ln$_2$O$_3$ powder and LnF$_3$ powder (99.99$\%$, Aladdin, Ln=Er, Dy, Ho), which are mixed up in a glove box with the molar ratios of Er$_2$O$_3$:ErF$_3$=0.998:1, Dy$_2$O$_3$:DyF$_3$=1:1, and Ho$_2$O$_3$:HoF$_3$=1:1, respectively. The mixtures are sealed in a quartz tube and slowly heated up to 800 $^{\circ}$C in a box furnace. After staying at 800 $^{\circ}$C for 60 minutes, the furnace is slowly cooled down to 600 $^{\circ}$C and then naturally cooled down to room temperature.
\par

The YbOCl single crystal is structurally analysed by a single crystal X-ray Diffractometer (XRD) (Bruker D8 Venture) and the structure was solved and refined using the Bruker SHELXTL Software Package. The polycrystals are measured using a powder XRD (Bruker D8 Da Vinci) with the mode of step scanning. The crystal structures are refined using the TOPAS program. Magnetic susceptibility, magnetization and heat capacity measurements from 1.8 K to 300 K are performed with a vibrating sample magnetometer (Quantum Design, Physical Property Measurement System, PPMS). For magnetic measurements, the polycrystalline samples which are pressed into thin plates, or YbOCl crystals are fixed on the sample holder with GE varnish. For heat capacity measurements, the samples are mounted with N grease for better thermal contact. The background heat capacity from the sample holder and grease is measured before sample is mounted, and subtracted in the analysis of heat capacity from sample.

\section{Results and discussions}

\begin{figure}
\centering
\includegraphics[width=\linewidth]{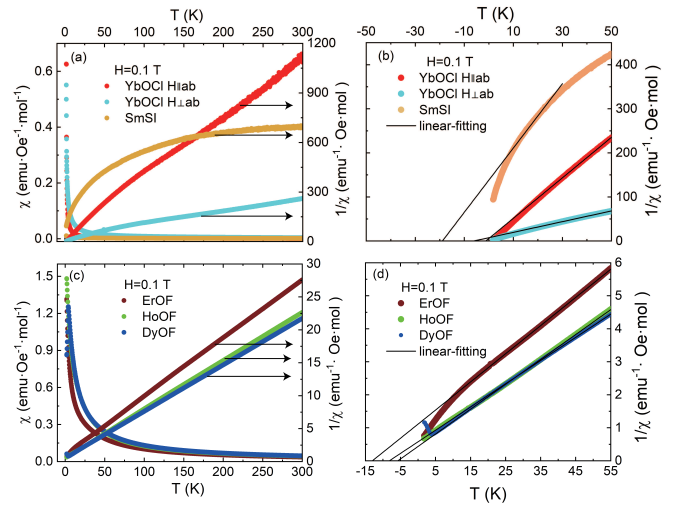}
\caption{Magnetic susceptibility for (a)$\&$(b) YbOCl/SmSI and (c)$\&$(d) ErOF/HoOF/DyOF. The temperature ranges for C-W fitting ((b)$\&$(d)) are estimated and listed in Table II and the factors for the estimations are discussed in the context. In DyOF, there exists a drop around 3 K, which suggests an AFM-like transition and is consistent with heat capacity results (Fig. 6). The other compounds show no obvious transition down to 1.8 K.}
\label{Fig.4}
\end{figure}
\par
We have determined the crystal structure of YbOCl and SmSI (Fig. \ref{Fig.1}). Both belong to the SmSI-type structure with a high symmetry of R-3m. A central rare-earth ion is coordinated by three halide anions and four chalcogen anions (see upper middle of Fig. \ref{Fig.1}). It is quite rare since such a rare-earth polyhedron contains two kinds rather than one kind of anions. Though the YbO$_4$Cl$_3$/SmS$_4$I$_3$ polyhedron looks a little more complicated, interestingly it retains the symmetry of C$_{3v}$. The configuration of CEF splitting is, in principle, determined by the point-group symmetries of crystals. This suggests that the the CEF splitting in YbOCl and SmSI is formally similar to the cases in YbMgGaO$_4$ and rare-earth chalcogenides \cite{Li2015,PhysRevLett.115.167203,Liu_2018,PhysRevB.103.035144}. In other words, one can still expect the configurations of four Kramers doublets for Yb$^{3+}$, eight Kramers doublets for Er$^{3+}$, three Kramers doublets for Sm$^{3+}$, etc. The basic structural feature is that two adjacent magnetic layers form a double-layer sheet which, as a whole, are well separated from the neighboring double-layer ones with a weak van der Waals inter-sheet coupling. The A/B sublattices of a bipartite honeycomb lattice are precisely assigned to two layers in a double-layer sheet. In other words, the spin sites in a double-layer sheet, although not exactly in a flat plane, form a regular honeycomb lattice without any bond distortion (see lower middle of Fig. \ref{Fig.1}). The spin configuration well matches the original Kitaev model. Furthermore, the nearest-neighbor Dzyaloshinskii-Moriya (DM) interaction is symmetrically prohibited and the next nearest-neighbor one is allowed. This is expected to bring more interesting spin physics. We will have more discussions on it later.
\par
\begin{figure}
\centering
\includegraphics[width=\linewidth]{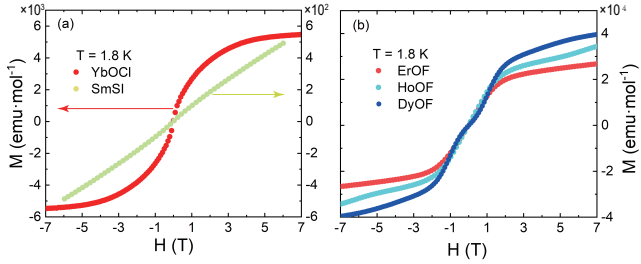}
\caption{Magnetization for (a) YbOCl/SmSI and (b) ErOF/HoOF/DyOF. SmSI reaches no magnetic saturation up to 7 T while the other four are saturated around 1$\sim$3 T.}
\label{Fig.5}
\end{figure}
\par

\begin{table}[htbp]
	\centering
	\caption{\label{distance} Some key distances in the refined crystal structures. a and c are lattice constants and the other ones are illustrated in Fig. \ref{Fig.1}and Fig. \ref{Fig.2}.}
	\begin{tabular}{cccccccc}
		\hline
		Comp.  & Sym. & a(nm) & c(nm) & d(nm) & D(nm) & d$_1$(nm) & d$_2$(nm)\tabularnewline
		YbOCl & R-3m & 0.37199 & 2.7791 & 0.28203 & 0.6443 & 0.3545 & 0.372 \tabularnewline
		SmSI & R-3m & 0.45575 & 3.26631 & 0.33464 & 0.75413 & 0.42570 & 0.45575 \tabularnewline
		ErOF & R-3m & 0.38741 & 1.88293 & 0.28068 & 0.34696 & 0.35569 & 0.37841 \tabularnewline
		HoOF & R-3m & 0.38015 & 1.89104 & 0.28189 & 0.34846 & 0.35726 & 0.38016 \tabularnewline
		DyOF & R-3m & 0.38197 & 1.90140 & 0.28344 & 0.35037 & 0.35912 & 0.38197 \tabularnewline \hline
	\end{tabular}
\end{table}
The structure of Er/Dy/HoOF is also determined using powder X-ray diffraction and Rietveld refinements (Fig. \ref{Fig.2}). Basically, it is isostructural to the SmSI-structure and the high symmetry of R-3m is still kept. The slight difference is the stacking order of the double-layer sheets, which is an A-B-C stacking in YbOCl/SmSI and becomes an A-C-B one in Er/Ho/DyOF (Fig. \ref{Fig.3}). Compared to YbOCl/SmSI, Er/Ho/DyOF series compounds exhibit larger ratios of intra-sheet to inter-sheet distances, which offers a potential way to manipulate the inter-layer coupling in both charge and spin channels. Some key distances have been listed in Table \ref{distance}.  We would like to highlight that the weak van der Waals coupling guarantees the true two-dimensionality of the family, which is a dispensable prerequisite for the exact realization of Abelian/non-Abelian anyons. In principle, even a small deviation from two dimensionality will invalidate the braiding statistics for anyons and hence is potentially disadvantageous to the scheme of topological computation.
\par
Before we go to magnetic properties of the compounds, it should be pointed out that the rare-earth ions with an odd number of 4f electrons generally exhibit an effective spin J$_{eff}$=1/2 related to Kramers doublets guaranteed by time reversal symmetry, while the ones with an even number of electrons having no Kramers doublets need to be discussed circumstantially. In our case, YbOCl, SmSI and ErOF with J$_{eff}$=1/2 should be more relevant to KSL, while Ho$^{3+}$ and Dy$^{3+}$ with larger moments may involve dipolar interactions and behave more classically.
\par

\begin{figure}
\centering
\includegraphics[width=\linewidth]{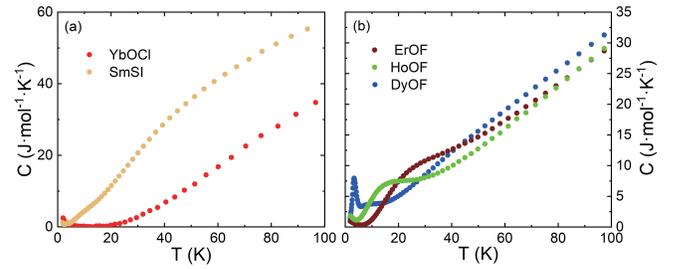}
\caption{Heat capacity measurements for (a) YbOCl/SmSI and (b) ErOF/HoOF/DyOF. The AFM-like transition in DyOF is also evidenced by the peak in heat capacity around 3 K (b). The broad features above 10 K in ErOF and HoOF may be related to rich CEF excitation levels in the two compounds. The origin of the upturn below 10 K in YbOCl (a) is not clear yet and needs to be identified by further experiments down to milikelvins.}
\label{Fig.6}
\end{figure}
\par

Fig. \ref{Fig.4} shows magnetic susceptibilities down to 1.8 K, which fall into two categories (Fig. \ref{Fig.4}a and \ref{Fig.4}c) according to the structural similarity. The Curie-Weiss (C-W) analysis has been made based on the susceptibilities (Fig. \ref{Fig.4}b and \ref{Fig.4}d). It should be noted that a reasonable temperature range for C-W analysis strongly depends on CEF excitations of rare-earth magnetic ions. More specifically, it requires that the upper limit is far below the first CEF excitation and the lower one is far above exchange coupling. Though the CEF excitations in the compounds are not investigated yet, generally a compound with better ionicity is expected to have higher CEF excitations. Therefore, one can presume that the compounds studied here have larger first CEF excitation levels comparable to those of other rare-earth oxides like YbMgGaO$_4$ \cite{PhysRevLett.118.107202}. Considering the exchange coupling of several kelvins for rare-earth magnetic ions due to the screening, we are able to estimate the effective C-W temperature ranges for the compounds. The C-W fitting details have been summarized in Table \ref{Fit}. It should be pointed out that the reverse susceptibility of SmSI in Fig. \ref{Fig.4}b shows a less linear behavior which outputs a relatively high C-W temperature of $\sim$18 K. At present it is unclear if the deviation from a linear form is related to the intrinsic spin and CEF excitations in SmSI, or extrinsically caused by the air-sensitivity. More careful investigations are required to clarify the issue in the future.
\par

\begin{table}[htbp]
	\centering
	\caption{\label{Fit} Magnetic parameters extracted from Curie-Weiss analysis $\chi$=C/(T-$\theta_{cw}$). $\mu_{eff}$ refers to the effective moments in units of Bohr magnetons. The temperature ranges for C-W fitting are listed here.}
	\begin{tabular}{cccccc}
		\hline
		Compounds  & Direction & C & $\theta_{cw}$(K) & $\mu_{eff}$($\mu_B$) &Temp. range (K) \tabularnewline
		YbOCl & c-axis & 0.21785 & -1.22 & 1.32 & 25-55  \tabularnewline
		YbOCl & ab-plane & 0.82372 & -6.15 & 2.567 & 25-55  \tabularnewline
		SmSI & / & 0.13716 & -18.94 & 1.0475 & 10-25  \tabularnewline
		ErOF & / & 11.72 & -13.0 & 9.683 & 25-55  \tabularnewline
		HoOF & / & 13.265 & -5.74 & 10.3 & 25-55  \tabularnewline
		DyOF & / & 14.2166 & -8.16 & 10.665 & 25-55  \tabularnewline \hline
	\end{tabular}
\end{table}

Magnetic susceptibilities demonstrate that all the compounds but DyOF show no obvious phase transitions at least down to 1.8 K. This is further evidenced by magnetization (Fig. \ref{Fig.5}) and heat capacity (Fig. \ref{Fig.6}) measurements. DyOF undergoes an AFM-like transition around 3 K. It remains to be uncovered that the transition is driven by spin-spin interactions or by dipolar interactions. YbOCl shows a magnetic anisotropy of $\sim$2:1 (Table \ref{Fit}) which is not very high and may favor the K term rather than anisotropic off-diagonal terms \cite{Sears2020}. And there is an unusual upturn in heat capacity of YbOCl with lowering temperatures down to 1.8 K (Fig. \ref{Fig.6}a). Further measurements down to milikelvins are highly required to figure out whether or not the upturn is an indication of a phase transition. ErOF and HoOF shows broad features in heat capacity at temperatures of above 10 K (Fig. \ref{Fig.6}b), which are higher than the C-W temperatures. The features are possibly related to a large number of CEF excitation levels in the two rare-earth ions and detailed discussions can be found somewhere. SmSI has a smaller moment but shows an unusual magnetization which is not saturated up to 7 T (Fig. \ref{Fig.5}a). As comparison, the other compounds are easily saturated around 1$\sim$3 T. And its magnetic susceptibility and the extracted C-W temperature are also remarkably distinct from those of the other ones. Whether or not this suggests a rare-earth spin system with much stronger spin interactions, needs to be explored in the future.
\par

As mentioned above, the strong SOC in rare-earth magnetic ions intrinsically provides highly anisotropic spin interactions, which have been well investigated in the triangular lattice QSL candidates \cite{PhysRevLett.115.167203}. When the anisotropic spin interactions are transplanted on an undistorted honeycomb lattice, we can fully expect the bond-dependent Kitaev interactions coming out of them. In fact, the generic nearest neighbor spin Hamiltonian of rare-earth ions on a honeycomb lattice can be equivalently written as J-K-$\Gamma$-$\Gamma^{'}$ model \cite{10.21468/SciPostPhysCore.3.1.004,PhysRevLett.112.077204}, in which the Kitaev term K is clearly highlighted. And the transformation relations of exchange parameters between the two models are derived \cite{10.21468/SciPostPhysCore.3.1.004}. As discussed above, the next nearest neighbor DM term is symmetrically allowed in our case. For accuracy, the DM term should be taken into account circumstantially and it may bring non-trivial effects in some cases. It has been predicted that the DM term in Kitaev materials can contribute to a non-quantized thermal Hall conductance under magnetic fields \cite{PhysRevResearch.1.013014}, which is understood in term of the motion of the emergent spinons twisted by an internal gauge flux.

\section{Summary}
In summary, we report the discovery of a rare-earth based family of KSL candidates. Most family members share the SmSI-type structure with the symmetry of R-3m and rare-earth magnetic ions form a regular honeycomb lattice. The combination of the honeycomb lattice and strongly spin-orbit coupled moments of 4f electrons, generates bond-dependent anisotropic exchange couplings required by Kitaev model. The family features a weak van der Waals interlayer coupling which is of particular advantage for exact anyonic excitations and potential applications. We have synthesized some representative family members including the crystals of YbOCl and the polycrystals of SmSI, ErOF, HoOF and DyOF. Their structures are carefully determined by Rietveld refinements. Thermodynamic measurements including magnetic susceptibility, magnetization and heat capacity, have been carried out down to 1.8 K, and no obvious phase transitions are found in the samples but DyOF. Thanks to the diversity of the family members, the family is expected to be a promising and manipulable platform for the study of KSL and may host more intriguing spin physics beyond Kitaev physics.

\subsection{ACKNOWLEDGEMENTS}
This work was supported by the National Key Research and Development Program of China (Grant Nos. 2017YFA0302904 and No. 2016YFA0300504), the National Science Foundation of China (Grant Nos. U1932215 and No. 11774419), and the Strategic Priority Research Program of the Chinese Academy of Sciences (Grant No. XDB33010100). Q.M.Z. acknowledges the support from Users with Excellence Program of Hefei Science Center and High Magnetic Field Facility, CAS. The refined structural data are available on request.

\bibliography{REChX_materials_v5.bib}

\end{document}